\documentclass[aps,twocolumn,english,superscriptaddress,citeautoscript,preprintnumbers,amsmath,amssymb,floatfix,footinbib]{revtex4-1}
\usepackage[frozencache=true,cachedir=minted-cache]{minted} 
\usepackage{graphicx}
\usepackage{graphics}               
\usepackage{subfigure}
\usepackage{bm}   
\usepackage{soul}   
\usepackage{amssymb } 
\usepackage{array}       
\usepackage[breaklinks=true,colorlinks,citecolor=blue,linkcolor=blue,urlcolor=blue]{hyperref}
\usepackage{float}
\usepackage[table]{xcolor}
\usepackage{cancel}
\usepackage[normalem]{ulem}

\footnotetext{These authors contributed equally to this work}
\definecolor{Cyan}{rgb}{0.4,1,0.7}
\definecolor{LightCyan}{rgb}{0.8,1,0.85}

\UseRawInputEncoding
\begin{document}
	
 \title{Single crystalline orthorhombic GdAlGe as a rare earth magnetic Dirac nodal-line metal}
	\author{Antu Laha\textsuperscript{**}}\email[]{antulaha.physics@gmail.com}
	\affiliation{Department of Physics and Astronomy, Stony Brook University, Stony Brook, New York 11794-3800, USA}
	\affiliation{Condensed Matter Physics and Materials Science Division, Brookhaven National Laboratory, Upton, New York 11973-5000, USA}
 \author{Juntao Yao \textsuperscript{**}}
		\affiliation{Condensed Matter Physics and Materials Science Division, Brookhaven National Laboratory, Upton, New York 11973-5000, USA}
       \affiliation{Department of Materials Science and chemical Engineering, Stony Brook University, Stony Brook, New York 11794-3800, USA}
        \author{Asish K. Kundu \textsuperscript{**}}\email[]{akundu@bnl.gov}
 \affiliation{National Synchrotron Light Source II, Brookhaven National Laboratory, Upton, New York 11973-5000, USA}
	\author{Niraj Aryal \textsuperscript{**}}\email[]{naryal@bnl.gov}
	\affiliation{Condensed Matter Physics and Materials Science Division, Brookhaven National Laboratory, Upton, New York 11973-5000, USA}
         
 \author{Anil Rajapitamahuni}
	\affiliation{National Synchrotron Light Source II, Brookhaven National Laboratory, Upton, New York 11973-5000, USA}
 \author{Elio Vescovo}
	\affiliation{National Synchrotron Light Source II, Brookhaven National Laboratory, Upton, New York 11973-5000, USA}
         \author{Fernando Camino}
	\affiliation{Center for Functional Nanomaterials, Brookhaven National Laboratory, Upton, New York 11973-5000, USA}
 \author{Kim Kisslinger}
	\affiliation{Center for Functional Nanomaterials, Brookhaven National Laboratory, Upton, New York 11973-5000, USA}
 \author{Lihua Zhang}
 \affiliation{Center for Functional Nanomaterials, Brookhaven National Laboratory, Upton, New York 11973-5000, USA}
 \author{Dmytro Nykypanchuk}
 \affiliation{Center for Functional Nanomaterials, Brookhaven National Laboratory, Upton, New York 11973-5000, USA}
\author{J. Sears}
	\affiliation{Condensed Matter Physics and Materials Science Division, Brookhaven National Laboratory, Upton, New York 11973-5000, USA} \author{J. M. Tranquada}
	\affiliation{Condensed Matter Physics and Materials Science Division, Brookhaven National Laboratory, Upton, New York 11973-5000, USA}
         \author{Weiguo Yin}
	\affiliation{Condensed Matter Physics and Materials Science Division, Brookhaven National Laboratory, Upton, New York 11973-5000, USA}
    	  \author{Qiang Li}\email[]{qiangli@bnl.gov}
	\affiliation{Department of Physics and Astronomy, Stony Brook University, Stony Brook, New York 11794-3800, USA}
	\affiliation{Condensed Matter Physics and Materials Science Division, Brookhaven National Laboratory, Upton, New York 11973-5000, USA}
\begin{abstract}
Crystal engineering is a method for discovering new quantum materials and phases, which may be achieved by external pressure or strain. Chemical pressure is unique in that it generates internal pressure perpetually to the lattice. As an example, GdAlSi from the rare-earth ($R$) $R$Al$X$ ($X =$ Si or Ge) family of Weyl semimetals is considered. Replacing Si with the larger isovalent element Ge creates sufficiently large chemical pressure to induce a structural transition from the tetragonal structure of GdAlSi, compatible with a Weyl semimetallic state, to an orthorhombic phase in GdAlGe, resulting in an inversion-symmetry-protected nodal-line metal. We find that GdAlGe hosts an antiferromagnetic ground state with two successive orderings, at $T_\mathrm{N1}$ = 35 K and $T_\mathrm{N2}$ = 30 K. In-plane isothermal magnetization shows a magnetic field induced metamagnetic transition at 6.2 T for 2 K. Furthermore, electron-hole compensation gives rise to a large magnetoresistance of $\sim 100\%$ at 2 K and 14 T. Angle-resolved photoemission spectroscopy measurements and density functional theory calculations reveal a Dirac-like linear band dispersion over an exceptionally large energy range of $\sim$ 1.5 eV with a high Fermi velocity of $\sim 10^6$ m/s, a rare feature not observed in any magnetic topological materials. 
\end{abstract}
	
\maketitle
	
\section{Introduction}
The interplay between topology and magnetism is a thriving research area in condensed matter physics. Theory predicts that the coupling between local magnetic moments in topological Weyl/Dirac semimetals can be mediated by conducting Weyl/Dirac fermions through the Ruderman-Kittel-Kasuya-Yosida (RKKY), Heisenberg, Kitaev, and Dzyaloshinskii-Moriya (DM) interactions, that can lead to rich spin textures and complex magnetism \cite{PhysRevB.104.024414, PhysRevB.103.155151, PhysRevB.96.115204, PhysRevB.92.241103}; however, experimental evidence of such couplings is limited. 

\begin{figure}
		\centering
		\includegraphics[scale=0.65]{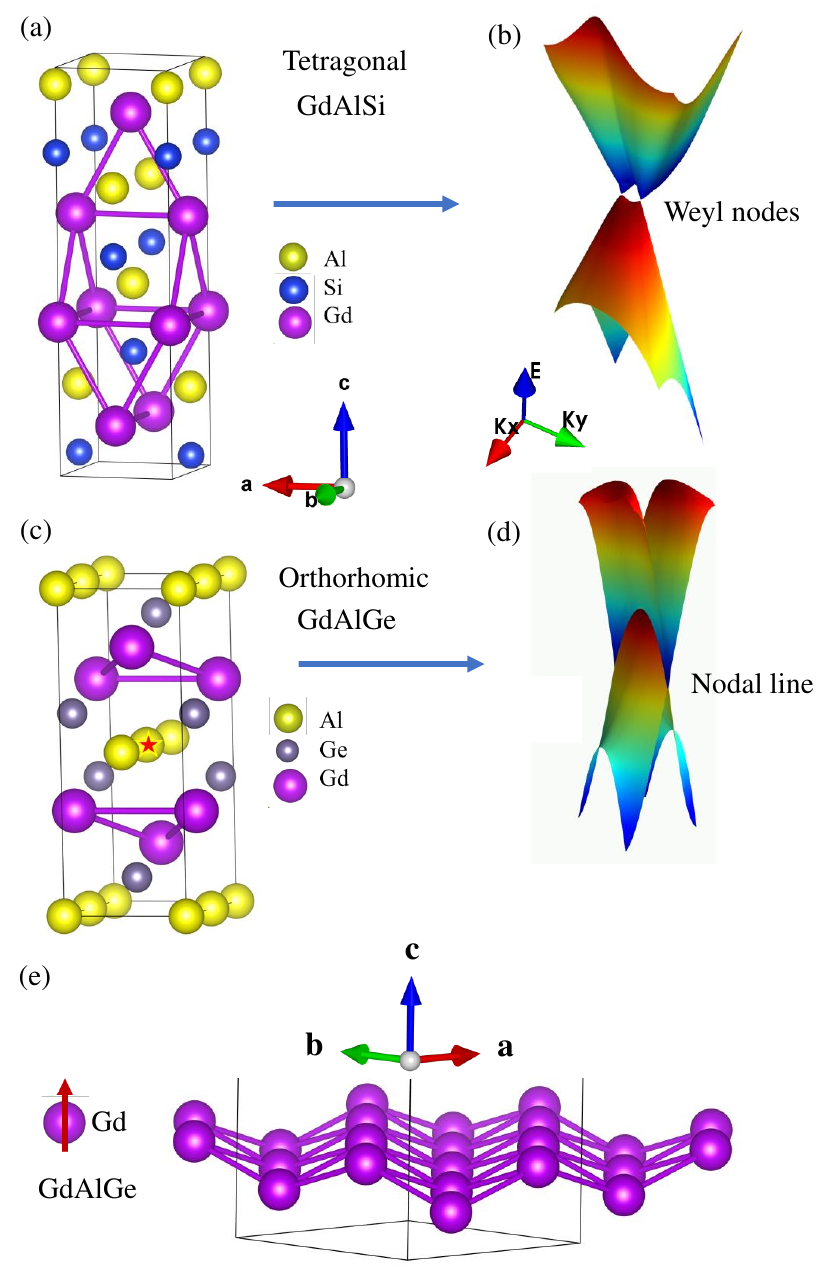} 
		\caption{A comparison of crystal structure and electronic structure of GdAlSi and GdAlGe. (a) Tetragonal crystal structure of GdAlSi. (b) Inversion symmetry broken tetragonal crystal structure gives rise to the Weyl nodes in the electronic structure \cite{GdAlSi_PRB_2024} (c) Orthorhombic crystal structure of GdAlGe. The star represents the point of inversion. (d) Inversion symmetry protected orthorhombic phase gives rise to nodal-lines in GdAlGe. (e) Buckled layer of Gd atoms in GdAlGe. The Gd atomic layer in GdAlSi has no such buckling structure. The arrow represents the spin configuration of Gd-atoms.}
		\label{Fig_Intro}
	\end{figure}

A relevant family for investigating such phenomena is $R$Al$X$ ($R$: Rare earth; $X$: Si or Ge).  The members of this family involving light rare-earth elements, such as La, Ce, Pr, Nd, and Sm, crystallize in a non-centrosymmetric LaPtSi-type tetragonal structure (space group $I4_1md$) illustrated in Fig.~\ref{Fig_Intro}(a).  These compounds are magnetic Weyl/Dirac semimetals exhibiting such complex magnetic behaviors including helical ferrimagnetic order, spiral magnetic order, and square-coordinated multi-${\bf k}$ magnetic phases \cite{NdAlSi_Nature_materials_2020, SmAlSi_PRX_2023, CeAlGe_PRL_2020, NdAlGe_PRB_2023}. In contrast, compounds involving heavy rare-earth elements, such as Dy, Ho, Tm, and Lu, crystallize in a distinct centrosymmetric orthorhombic structure (space group $Cmcm$). \cite{PUKAS2004162, YAlSi_MDPI_2022, TOBASH200658, BOBEV20052091, WANG2022163623}. The topological properties and electronic structure of the orthorhombic phase remains less explored for lack of single crystals.

GdAl$X$ is special, since Gd ($4f^7$) sits in the middle of the rare earth series that runs from La ($4f^0$) to Lu ($4f^{14}$).  Positioned between the compounds that adopt the non-centrosymmetric (light $R$) or centrosymmetric (heavy $R$) phases of $R$Al$X$ regardless the choice Si or Ge for $X$, it is the optimal situation for chemical pressure to have an impact.  GdAlSi is known to be tetragonal \cite{GdAlSi_PRB_2024}, and a metastable tetragonal phase has been demonstrated in GdAlGe polycrystalline samples obtained by quenching from high temperature \cite{GdAlge_IOP_2020}.  Here we demonstrate that single crystals of orthorhombic GdAlGe can be obtained by slow cooling.  This gives us the opportunity to explore the properties of the centrosymmetric phase.

As one might expect and as we will show, orthorhombic GdAlGe has a different electronic band structure compared to that of GdAlSi. Instead of the Weyl nodes observed in GdAlSi [Fig.\ref{Fig_Intro}(b)] \cite{GdAlSi_PRB_2024}, inversion-symmetry-protected nodal lines are found in GdAlGe, as shown in Fig.~\ref{Fig_Intro}(d). Moreover, GdAlGe has a different magnetic ground state from that of GdAlSi, leading to a new magnetic field versus temperature phase diagram. Two successive antiferromagnetic orderings (at  $T_\mathrm{N1}$ = 35 K and $T_\mathrm{N2}$ = 30 K) are observed in the in-plane magnetic susceptibility of GdAlGe, and in-plane magnetization shows a metamagnetic transition at 6.2 T for $T=2$~K. The ordering temperature $T_\mathrm{N1}$ = 35 K is the highest in the $R$Al$X$ family. Our results from angle-resolved photoelectron spectroscopy (ARPES) show Dirac-like linear band dispersion for binding energies extending from the Fermi energy to 1.5 eV. Such a large energy range of Dirac dispersion is rare, and not observed in any magnetic topological materials.

\section{Methods}
\subsection{Experimental details}
\begin{figure*}[htp]
		\centering
		\includegraphics[scale=0.6]{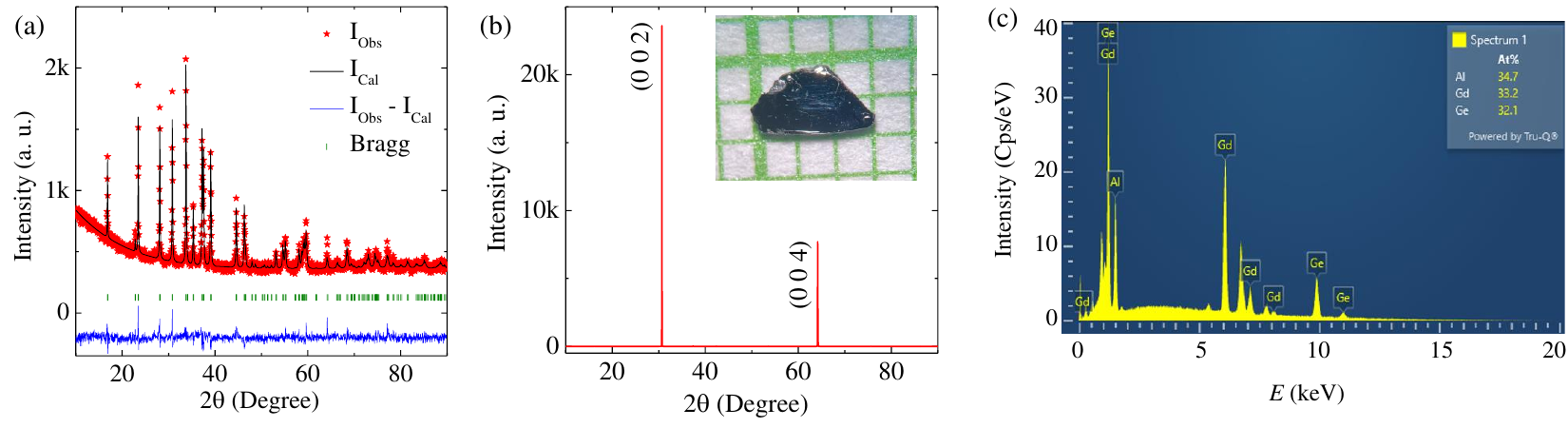} 
		\caption{(a) Refined powder x-ray diffraction pattern of crushed GdAlGe single crystals, recorded at room temperature. The observed intensity (red scattered points), Rietveld refinement fit (solid black line), difference between the experimentally observed and calculated intensities (solid blue line), and Bragg peak positions (vertical green bars) are shown. (b) X-ray diffraction pattern of single crystal. An image of a single crystal is shown in the inset (the grit scale: 1 mm). (c) Energy-dispersive X-ray spectroscopy confirms the stoichiometry of the compound. }
		\label{Fig1}
	\end{figure*}

GdAlGe single crystals were grown by the standard self flux method with excess Al as a flux \cite{Z_Fisk_Al_flux_growth, GdAlSi_PRB_2024}. Gd ingot (99.9$\%$), Al ingot (99.999$\%$), and Ge chips (99.999$\%$) in a molar ratio of 1:10:1 were mixed in an alumina crucible. The crucible was then sealed into a quartz tube under a partial pressure of argon gas. The content was heated to 1100$^\circ$C, kept for 24 hours at that temperature, and then cooled to 700$^\circ$C at a rate of 2$^\circ$C/hour. Plate-like single crystals were extracted from the flux by centrifuging. The typical size of the crystal is 3mm$\times$2mm$\times$0.5mm as shown in the inset of Fig.~\ref{Fig1}(b). The crystal structure and phase purity were investigated by x-ray diffraction technique using Cu-K$_\alpha$ radiation in the Rigaku Smartlab diffractometer. The chemical compositions were confirmed by energy dispersive x-ray spectroscopy (EDS) measurements in a a JEOL JSM-7600F scanning electron microscope. Magnetotransport measurements were carried out in a physical property measurement system (PPMS, Quantum Design) via the standard six-probe method and magnetic measurements were carried out in a magnetic property measurement system (MPMS, Quantum Design).
	
ARPES experiments were performed at the Electron Spectro Microscopy (ESM) 21-ID-1 beamline of the National Synchrotron Light Source II, located at Brookhaven National Lab \cite{rajapitamahuni2024electron}. The beamline is equipped with a Scienta DA30 electron analyzer, with base pressure $\sim$ 1$\times$10$^{-11}$ mbar. Prior to the ARPES experiments, samples were cleaved using a post inside an ultra-high vacuum chamber (UHV) at $\sim$ 18 K. The total energy and angular resolution during the ARPES measurements were $\sim$ 20 meV and $\sim$ 0.2$^\circ$, respectively. 

\subsection{Computational details}
The density-functional-theory (DFT) calculations were done using the VASP DFT package~\cite{VASP1,VASP2,PAW}. The Brillouin zone was sampled with a regular mesh of $11 \times 11 \times 5$ $k$-points. 
In order to resolve the small energy differences between various magnetic phases, we used a  \textit{k}-mesh of up to $21 \times 21 \times 13$.
Perdew-Burke-Ernzerhof (PBE) exchange-correlation functional~\cite{PBE} within the generalized gradient approximation (GGA) was used in all the calculations.  The GGA + $U_{\textrm{eff}}$ method ~\cite{LDAU_Dudarev} was used to handle the Gd-4\textit{f} orbitals.
$U_{\textrm{eff}}$ of 6 eV~\cite{Eu_aryalPRB2022,GdAlSi_PRB_2024} was chosen in our calculations. The calculations for the non-magnetic phase were done within open-core approximation by freezing the Gd-4\textit{f} states. The spin-orbit coupling (SOC) was treated in the second variational method as implemented in VASP.
 
\section{Results}
\subsection{Crystal structure and sample characterization}
Fig.~\ref{Fig1}(a) shows the Rietveld refinement of powder x-ray diffraction data using the FULLPROF software package \cite{Fullprof_1993}. The refined pattern confirms that GdAlGe crystallizes in a YAlGe-type orthorhombic structure with space group $Cmcm$ (No. 63). 
The obtained lattice parameters from the XRD refinement fitting are $a$= 4.06367(19) \AA $~$, $b$ = 5.8035(2) and $c$ = 10.5249(5) \AA. No impurity phase was observed within our experimental resolution. Figure~\ref{Fig1}(b) shows the x-ray diffraction pattern for the $c$-axis of a crystal, which confirms that the plate-like surface is $ab$-plane. The EDS spectrum confirms the single-phase nature and correct stoichiometry of the grown crystals as shown in Fig.~\ref{Fig1}(c).  

	\begin{figure*}
		\centering
		\includegraphics[scale=0.75]{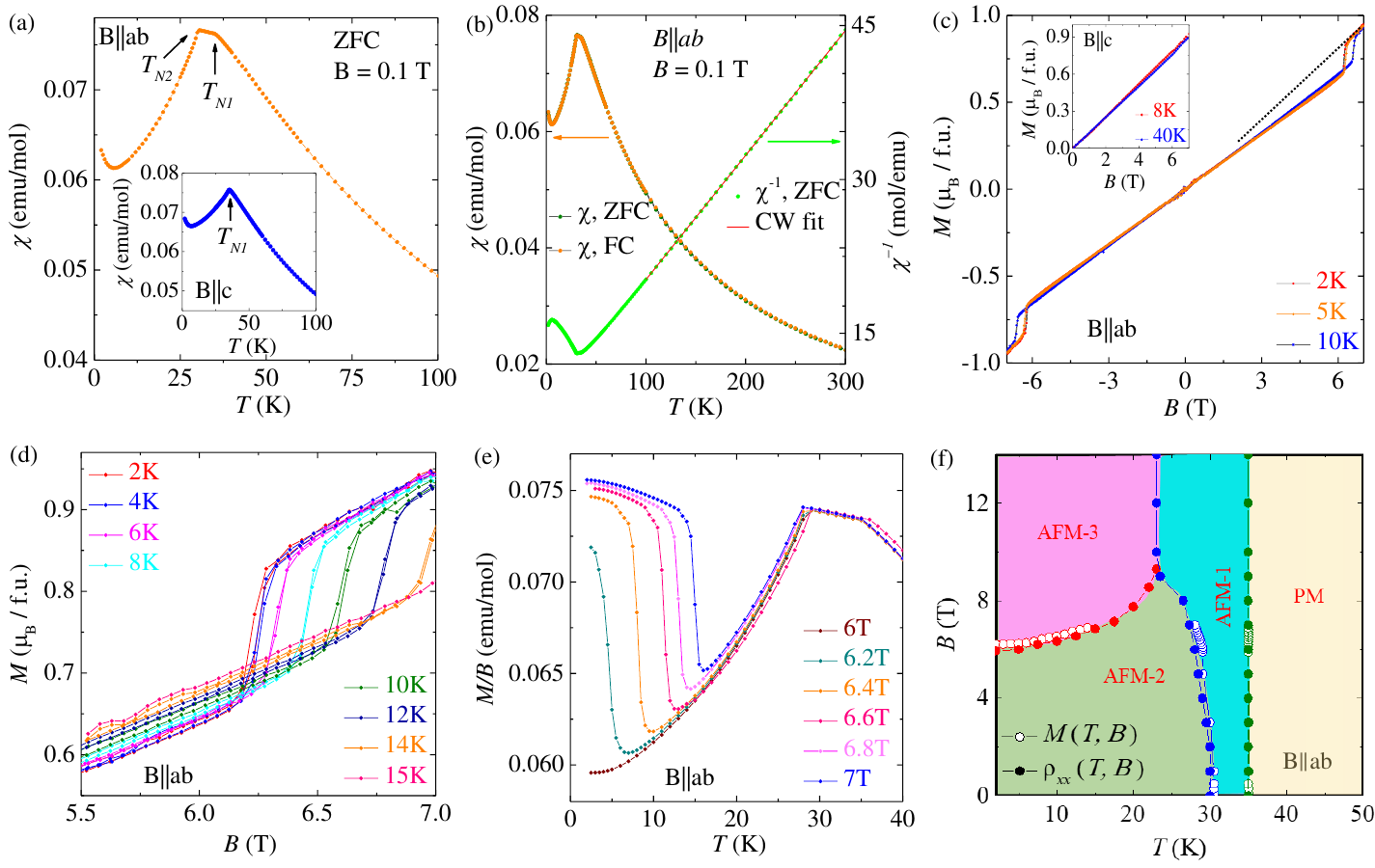} 
		\caption{(a) Temperature-dependent magnetic susceptibility in zero-field-cooled (ZFC) condition for $B\parallel ab$, $T_{N1}$ and $T_{N2}$ are two antiferromagnetic transition temperatures. Only one antiferromagnetic transition at $T_{N1}$ is observed for $B \parallel c$ as shown in the inset. (b) Temperature-dependent in-plane magnetic susceptibilities in zero-field-cooled (ZFC) and field-cooled (FC) condition overlap with each other. The inverse magnetic susceptibility is fitted with Curie-Weiss law in the temperature range 100 K $-$ 300 K. (c) Isothermal magnetization $M(B)$ as a function of magnetic field for $B\parallel ab$, The slope of the $M(B)$ curve is changed after the metamagnetic transition as indicted by dotted straight line. The inset shows $M(B)$ vs. $B$ for $B\parallel c$. (d) Isothermal $M(B)$ curves show metamagnetic transition for $B_c \ge 6.2$ T, $B_c$ is defined as the starting point for the sudden jump of $M(B)$. (e) Temperature-dependent magnetic susceptibility at various magnetic fields from 6 T to 7 T. (f) Temperature-magnetic field phase diagram for $B\parallel ab$. Three antiferromagnetic regions are denoted by AFM-1, AFM-2, and AFM-3; paramagnetic region is denoted by PM. AFM-1 and AFM-2 regions are constructed by observing the field dependence behavior of $T_{N1}$ and $T_{N2}$, AFM-3 regions is constructed from the temperature dependence of $B_c$. The data taken from magnetization ($M(T,B)$) and electrical resistivity ($\rho_{xx}(T,B)$) are denoted by hollow circle and solid circle respectively.} 
		\label{Fig2}
	\end{figure*}
 
\subsection{Magnetic properties}
Magnetic susceptibility shows two successive antiferromagnetic transitions at $T_{N1}$ = 35 K and $T_{N2}$ = 30 K for $B\parallel ab$, whereas the second transition at $T_{N2}$ is not observed for $B \parallel c$ as shown in Fig.\ref{Fig2}(a). Such a two step magnetic ordering was previously found in other rare-earth based ternary compounds \cite{NdAlSi_Nature_materials_2020, Ce2Ni3Ge5_JMMM_2018,SmAlSi_PRX_2023}. We note that only one antiferromagnetic transition at 34 K was observed in polycrystalline GdAlGe \cite{GdAlGe_JMMM_2021}. This comparison suggests that the two-step antiferromagnetic ordering is sensitive to the orientation of the applied field, even though GdAlGe exhibits a very weak magnetocrystalline anisotropy ($\chi_{c}/\chi_{ab}= 1.08$). 

No difference is observed between zero-field cooled (ZFC) and field cooled (FC) magnetic susceptibility for the temperature range 2 K -- 300 K as shown in Fig.~\ref{Fig2}(b). The effective magnetic moment of the Gd ions is estimated by fitting the inverse susceptibility ($1/\chi$) with the modified Curie-Weiss law, $\chi(T)=\chi_0+C/(T-\theta_p)$, in the paramagnetic region 100 K -- 300 K. Here, $\chi_0$, $C$, and $\theta_P$ are the temperature-independent susceptibility, Curie constant, and paramagnetic Curie temperature, respectively. The fitted effective magnetic moment of Gd$^{3+}$ is 8.16$\mu_B$, which is close to the theoretical value of $g\sqrt{S(S+1)}\mu_B=7.94\mu_B$ for $S=7/2$. The fitted value of the $\theta_P$ is negative ($\sim -68 $K), which is consistent with an AFM ground state in this compound. The estimated frustration parameter $f = - \theta_P / T_N \sim 2$ is moderate compared to the strongly geometrically frustrated magnets ($f > 10$) \cite{annurev:/content/journals/10.1146/annurev.ms.24.080194.002321}. 

Isothermal magnetization $M(B)$ shows linear field dependence behavior without saturation as shown in Fig.~\ref{Fig2}(c). Interestingly, we observe a sharp increase in magnetization around a critical field of $B_c \sim $ 6.2 T at 2 K for $B\parallel ab$ [Fig.~\ref{Fig2}(d)], which indicates a field induced metamagnetic transition. The slope of the $M(B)$ curve is changed after the metamagnetic transition as shown in Fig.~\ref{Fig2}(c), which indicates a canted spin state above $B_c$. No such transition is observed in the $M(B)$ for $B\parallel c$ up to our field limit of 7~T (the highest applied magnetic field in our MPMS). This transition was missed in previous studies of polycrystalline samples where the highest applied field was limited to 5 T \cite{GdAlGe_JMMM_2021}. To obtain further insight, we have measured the field dependence of magnetization at several temperatures from 2 K to 20~K, and the temperature dependence of magnetization at several magnetic fields from 6~T to 7~T  [Fig.~\ref{Fig2}(d), \ref{Fig2}(e)]. The critical field $B_c$ gradually increases with increasing temperature, and it reaches 7~T at around 14~K. The temperature dependence of $B_c$ is consistent with that obtained from the sharp drop in the temperature dependent magnetization as shown in Fig.~\ref{Fig2}(e).  For better visualization of the complex magnetism, we construct a temperature-magnetic field phase diagram as shown in Fig.~\ref{Fig2}(f). Enhancement of $B_c$ upon increasing temperature is widely seen in other metamagnetic systems \cite{CuMnAs_JMMM_2019, PhysRevLett.116.097204, 10.1063/1.3213100}, which can be interpreted using the molecular field theory \cite{doi:10.1080/00018730110101412}.  Alternatively, the presence of two magnetic transitions at zero field plus the metamagnetic transition shows a similarity to SmAlSi, where there is evidence of spiral magnetic order \cite{SmAlSi_PRX_2023}.

	\begin{figure*}
		\centering
		\includegraphics[scale=0.75]{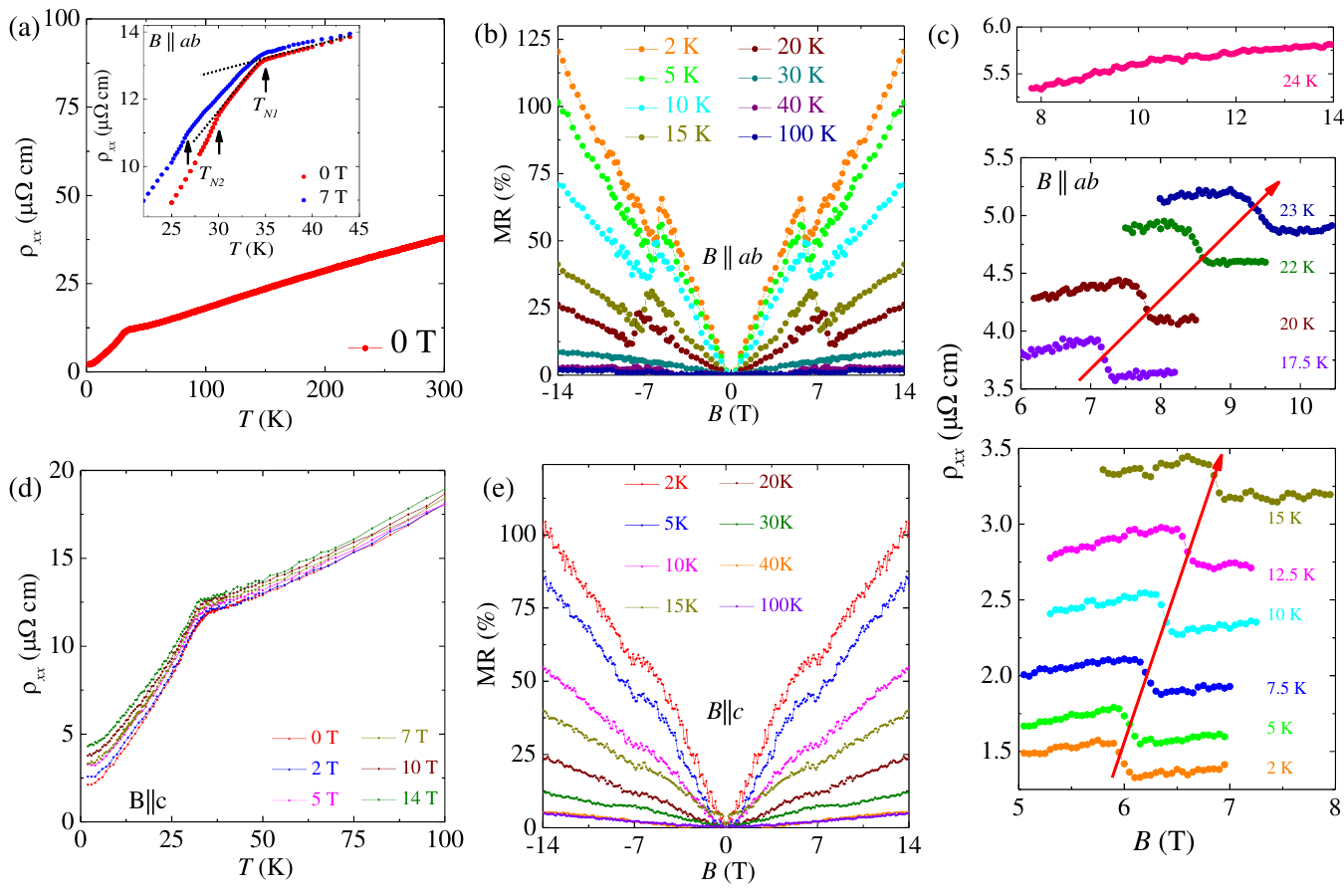} 
		\caption{(a) Electrical resistivity as a function of temperature. The residual resistivity ratio (RRR) = $\rho_{xx}$(300 K)/$\rho_{xx}$(2 K) = 18. Two antiferromagnetic transitions ($T_{N1}$ and $T_{N2}$) are demonstrated in the inset. (b) Magnetic field dependence of magnetoresistance (MR) at various temperatures for $B\parallel ab$, MR = $[(\rho_{xx}(B) - \rho_{xx}(0))/\rho_{xx}(0)] \times 100 \%$. (c) The resistivity jumps due to the metamagnetic transition is observed from 2 K to 23 K for $B\parallel ab$. (d) Electrical resistivity as a function of temperature at various magnetic field for $B\parallel c$.  (e) Magnetic field dependence of magnetoresistance (MR) at various temperatures for $B\parallel c$.}

		\label{Fig3}
	\end{figure*}
 
\subsection{Electrical resistivity and magnetoresistance}
The in-plane electrical resistivity, $\rho_{xx}$, decreases with decreasing temperature, and reaches 2.1 $\mu \Omega$~cm at 2~K [Fig.~\ref{Fig3}(a)]. This leads to a residual resistivity ratio (RRR) of $\sim$ 18, which is comparatively larger than that observed in other members of this family such as GdAlSi (RRR $\sim$ 3), NdAlSi (RRR $\sim$ 6), NdAlGe (RRR $\sim$ 2), and SmAlSi (RRR $\sim$ 4.95) \cite{GdAlSi_PRB_2024, NdAlSi_Nature_materials_2020, NdAlGe_PRM_2023, SmAlSi_IOP_2022}. Under zero field, the temperature-dependent resistivity curve exhibits a slope change at 35 K, followed by a pronounced decrease below 30 K. These features correspond to two antiferromagnetic transitions ($T_{N1}$ and $T_{N2}$) as confirmed by the magnetic susceptibility measurement. The $\rho_{xx}$ vs $T$ data series were taken at fixed magnetic fields for $B\parallel ab$, with the data at 0 and 7 T shown in the inset of Fig.~\ref{Fig3}(a) for clarity. The transition temperature $T_{N1}$ is found to be independent of the magnetic field, while $T_{N2}$ decreases with increasing $B$, consistent with the observations from magnetic susceptibility measurement for $B \parallel ab$. The field dependent behavior of $T_{N1}$ and $T_{N2}$ is illustrated in the $B\sim T$ phase diagram [Fig.~\ref{Fig2}(f)].

The $\rho_{xx}$ measurement for $B \parallel ab$ [Fig.~\ref{Fig3}(b)] exhibits a sharp jump at a critical field $B_c$ = 6 T at $T$ = 2 K, attributed to the metamagnetic transition as shown by the magnetization measurements discussed earlier. In Fig.~\ref{Fig3}(c), we present a series of resistivity data as a function of field near $B_c$ at various temperatures. We observed the $B_c$ increases with increasing temperature up to 23 K. At 24 K, the metamagnetic transition is suppressed with no jumps detected. The temperature dependence of $B_c(T)$ derived from $\rho_{xx} \sim B$ measurements is consistent with that observed in $M \sim B$ measurements. The $B_c(T)$ values determined from resistivity (solid circles) are plotted alongside those from magnetization (hollow circles) in the $B\sim T$ phase diagram [Fig.~\ref{Fig2}(f)]. Additionally, a substantial non-saturating magnetoresistance of $\sim 120 \%$ for $B \parallel ab$ and $\sim 100 \%$ for $B \parallel c$ are observed at 2 K and 14 T [Fig.~\ref{Fig3}(b) and ~\ref{Fig3}(e)]. Such a non-saturating MR has been reported in several topological materials having linearly dispersive bands and charge carrier compensations. \cite{Cd3As2_natmat_2015, NbP_natphy_2015, ZrSiS_PNAS_2017, WTe2_2014_nature, CaCdSn_PRB_2020, YbCdGe_PRB_2019, YbCdSn_PRB_2020}. We note that the electron-hole compensation effect may contribute to the large MR in GdAlGe, as the carrier densities of the hole and electron pockets are almost the same [see Fig.~\ref{Fig5}(c) and \ref{Fig5}(d)].

We also measured the temperature dependence of $\rho_{xx}$ for $B \parallel c $ under various magnetic fields, as shown in Fig.~\ref{Fig3}(d). The kink at 35 K ($T_{N1}$), attributed to the antiferromagnetic (AFM) ordering, remains unaffected by the application of magnetic fields up to $B = 14$ T, indicating a robust long range AFM order. The field dependence of magnetoresistance for $B \parallel c $ [Fig.~\ref{Fig3}(e)] reveals a small anomaly (a subtle kink) at $\sim 7 $~T for 2 K, which might be related to the influence of the metamagnetic transition. This anomaly is evident at temperatures up to $T_{N1}$, and appears to be temperature-independent. It is worth noting that in magnetization measurements for $B \parallel c $, the potential anomaly at 7 T could not be directly examined, as 7~T is at upper limit of the magnetization measurement range.

	\begin{figure}
		\centering
		\includegraphics[scale=0.5]{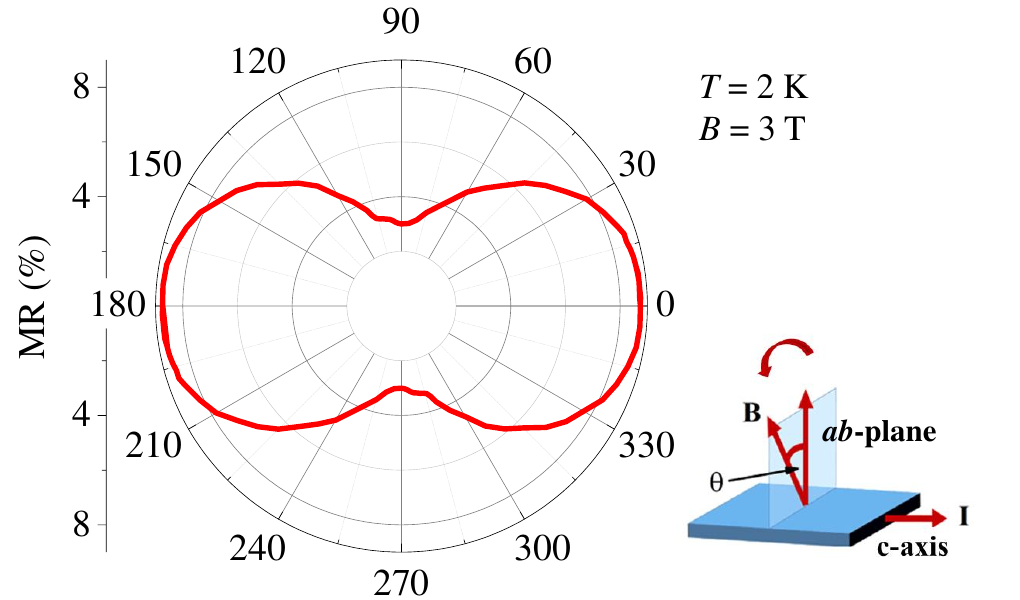} 
		\caption{In-plane anisotropy: The resistivity is measured along c-axis ( $\rho_{zz}$) by rotating the magnetic field in the $ab$-plane as shown in the schematic diagram. The magnetoresistance (MR) as a function of rotating angle $\theta$ for $T$ = 2 K and $B$ = 3 T.}
		\label{Anisotropy}
	\end{figure}
    
Another intriguing property of orthorhombic crystals is their in-plane anisotropy. To investigate this, we measured the resistivity along c-axis ( $\rho_{zz}$) by rotating the magnetic field within the $ab$-plane, as illustrated in the schematic diagram [Fig.\ref{Anisotropy}]. The magnetoresistance (MR) exhibits a two-fold symmetric pattern, demonstrating significant in-plane anisotropy. For instance, at $T = 2 $ K and $B = 3$ T, the MR reaches its maximum at 0$^\circ$ ($B\parallel a$) and minimum at 90$^\circ$ ($B\parallel b$), as shown in Fig.\ref{Anisotropy}. The in-plane magnetic anisotropy in GdAlGe holds potential for applications in magnetic sensing and spin-hall-effect-based MR devices.

\begin{figure}
		\centering
		\includegraphics[scale=0.35]{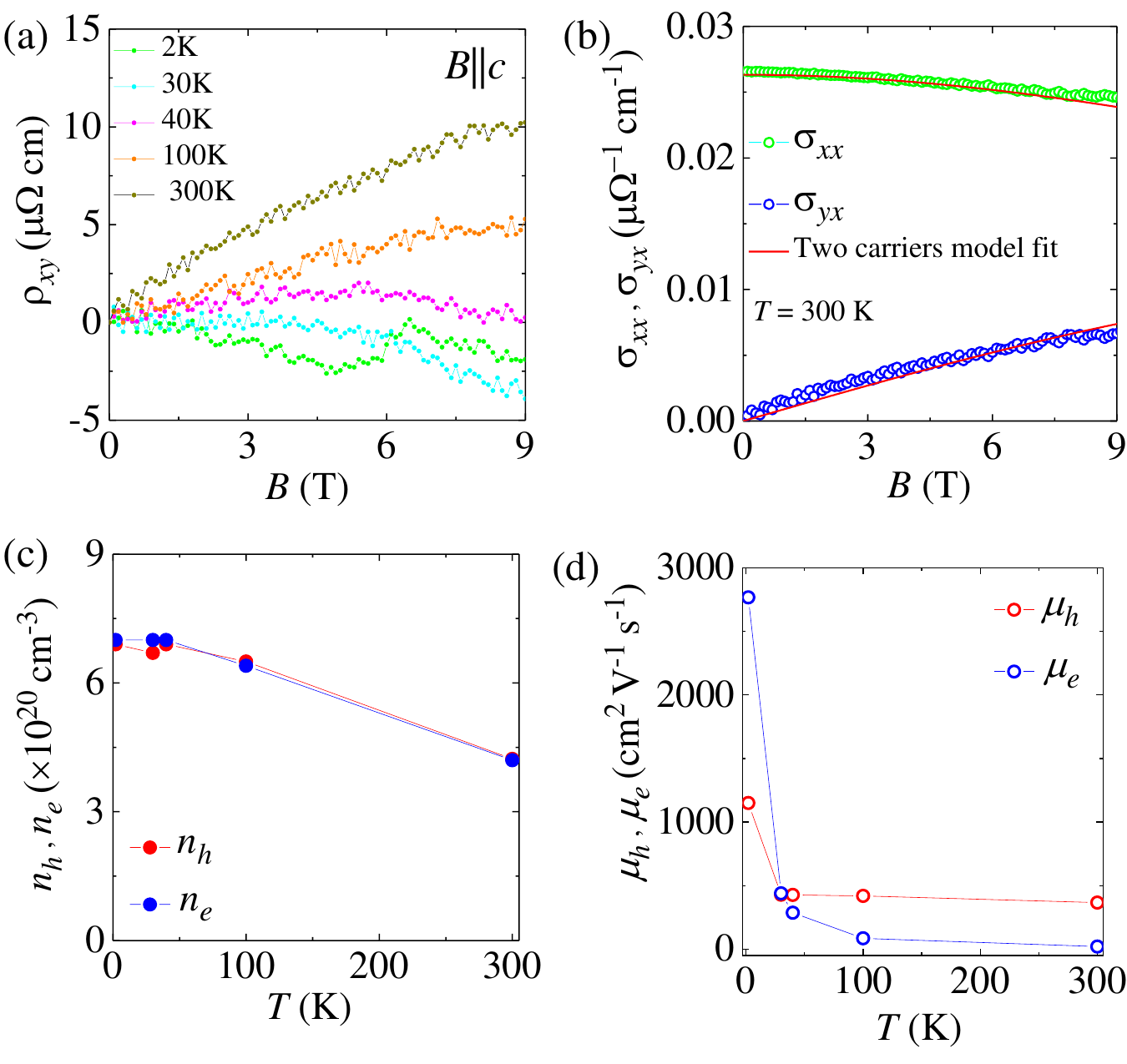} 
		\caption{(a) Hall resistivity ($\rho_{xy}$) as function of $B$ at various temperatures. (b) Simultaneous fitting of longitudinal conductivity ($\sigma_{xx}$) and Hall conductivity ($\sigma_{yx}$) with two carries model (equ. \ref{yy} and equ. \ref{yx}). (c) Hole (electron) density $n_h$ ($n_e$), and (d) mobility $\mu_h$ ($\mu_e$) as a function of temperature.}
		\label{Hall}
	\end{figure}

To determine the charge carriers densities and mobilities of GdAlGe, we measured magnetic field dependence of Hall resistivity ($\rho_{xy}$) and longitudinal resistivity ($\rho_{xx}$) at various temperatures from 2 K to 300K. The field dependence of $\rho_{xy}$ is not linear [Fig.\ref{Hall}(a)], which indicates that both electrons and holes contribute to the transport. The longitudinal conductivity ($\sigma_{xx}$) and Hall conductivity ($\sigma_{yx}$) are obtained using the expressions $\sigma_{xx} = \frac{\rho_{xx}}{\rho_{xx}^2 + \rho_{xy}^2}$, and $\sigma_{yx} = \frac{\rho_{xy}}{\rho_{xx}^2 + \rho_{xy}^2}$. The $\sigma_{xx}$ and  $\sigma_{yx}$ are simultaneously fitted with the semiclassical two carriers model \cite{Two_band_model} as shown in Fig.\ref{Hall}(b),

\begin{equation}
  \sigma_{xx} = e \left[\frac{n_h \mu_h}{1 + \mu_h^2 B^2} + \frac{n_e \mu_e}{1 + \mu_e^2 B^2}\right] 
  \label{yy}
\end{equation}

\begin{equation}
     \sigma_{yx} = e B \left[\frac{n_h \mu_h^2}{1 + \mu_h^2 B^2} - \frac{n_e \mu_e^2}{1 + \mu_e^2 B^2} \right]
     \label{yx}
\end{equation}

where $n_h$ ($n_e$) and $\mu_h$ ($\mu_e$) are the hole (electron) density and mobility, respectively.  The obtained carrier densities (error bar: $\pm 7\%$) and mobilities (error bar: $\pm 7\%$) from the two carriers model fitting are plotted as a function of temperature in Fig.\ref{Hall}(c) and \ref{Hall}(d). The ratio $n_h/n_e$ is $\sim 1 \pm 7\%$, which indicates the occurrence of electron-hole compensation in this compound.

\subsection{Electronic band structure}

\begin{figure}[b]
		  \includegraphics[width=0.48\textwidth]{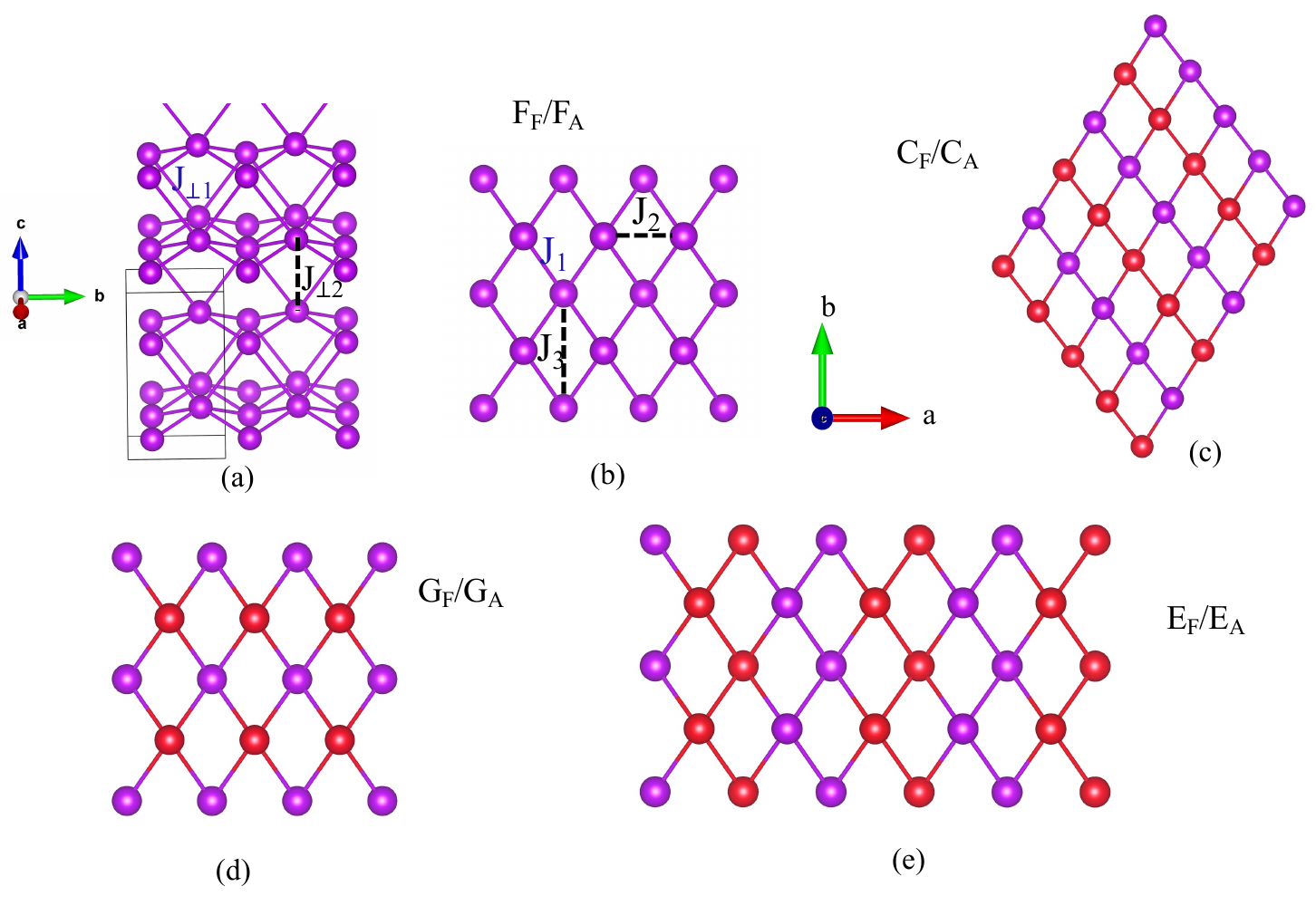}
		\caption{(a) Side view of crystal structure with only Gd atoms showing in-plane and out-of-the-plane nearest neighbour atomic coordination and corresponding exchange coupling parameters.
  (b-e ) Top view of a single Gd-atomic layer showing spin configurations for different magnetic patterns studied in this work using DFT calculations. The red and purple colors  represent up and down  spin configurations of Gd-atoms. Both  ferromagnetic and anti-ferromagnetic configurations in out-of-the-plane direction has been considered for each of the magnetic structures shown in the figure.  See text for details.}
		\label{fig:magnetic_patterns}
	\end{figure}



In Fig.~\ref{fig:magnetic_patterns}, we show different magnetic patterns studied in this work using DFT calculation. 
For each of the patterns shown in Fig.~\ref{fig:magnetic_patterns}, we studied both ferromagnetic (FM) as well as anti-ferromagnetic (AFM) patterns in out-of-the-plane direction, which are indicated by F and A subscripts.
For example, F$_F$ and F$_A$ patterns both have in-plane FM spin configuration; however, the spins linked by the shortest out-of-the-plane Gd-Gd bonds align ferromagnetically and antiferromagnetically, respectively. The G-type patterns have nearest neighbour (nn) AFM configuration in the plane, whereas C-type patterns have AFM interaction with second and third nearest neighbour and requires unit cell of size $\sqrt 2 \times \sqrt 2$. Similary, E-type patterns have FM interaction with nearest neighbours running along ${\textbf b}$-direction in zig-zag manner and AFM interaction with nnn  along ${\textbf a}$-direction.
Out of these magnetic configurations, we found E$_F$ configuration to be the lowest in energy, whereas F$_A$ and C-type configurations are competing in energy as shown in  Table.~\ref{table:MagneticPatternEnergy}. By mapping the results from our DFT energy calculations (Table I) to a simplified Heisenberg Hamiltonian, we found that while the first and third nearest neighbors of Gd atoms have weak ferromagnetic interactions, the second nearest neighbors have a strong anti-ferromagnetic interaction. Similarly, nearest neighbor out-of-plane coupling is ferromagnetic, whereas the next-nearest coupling in the out-of-plane direction is strongly antiferromagnetic. Such constraint is only satisfied by E-type magnetic pattern with out-of-plane ferromagnetic interaction. This analysis is indeed consistent with the DFT calculations.
The fact that the calculated energy differences between various AFM configurations is so small, comparable to the N\'eel temperature, is reminiscent of the rich magnetic correlations in iron-based superconductors~\cite{Unified_YinPRL2010} and suggests that a spiral character to the order is not ruled out.
From the DFT energy calculations, we estimated the Heisenberg exchange interactions to be as follows: $J_1S^2$=-0.16, $J_2S^2$=1.8, $J_3S^2$=-0.23, $J_{\perp 1}S^2$=-0.26,$J_{\perp 2}S^2$=0.84 and $J_{\perp 3}S^2$= 0.74 meV, where $J_1$ ($J_{\perp 1}$), $J_2$ ($J_{\perp 2}$), and $J_3$ ($J_{\perp 3}$) denote three nearest neighbor exchange coupling interactions along in-plane (out-of-plane) directions. From these values of exchange coupling parameters, we estimated N\'eel  temperature ($T_N$) to be $\sim$30 K ~\cite{HeisenbergModel_Pires} which is remarkably close to the experimental value of 35 K.

\begin{figure*}
\centering
  \includegraphics[width=0.9\textwidth]{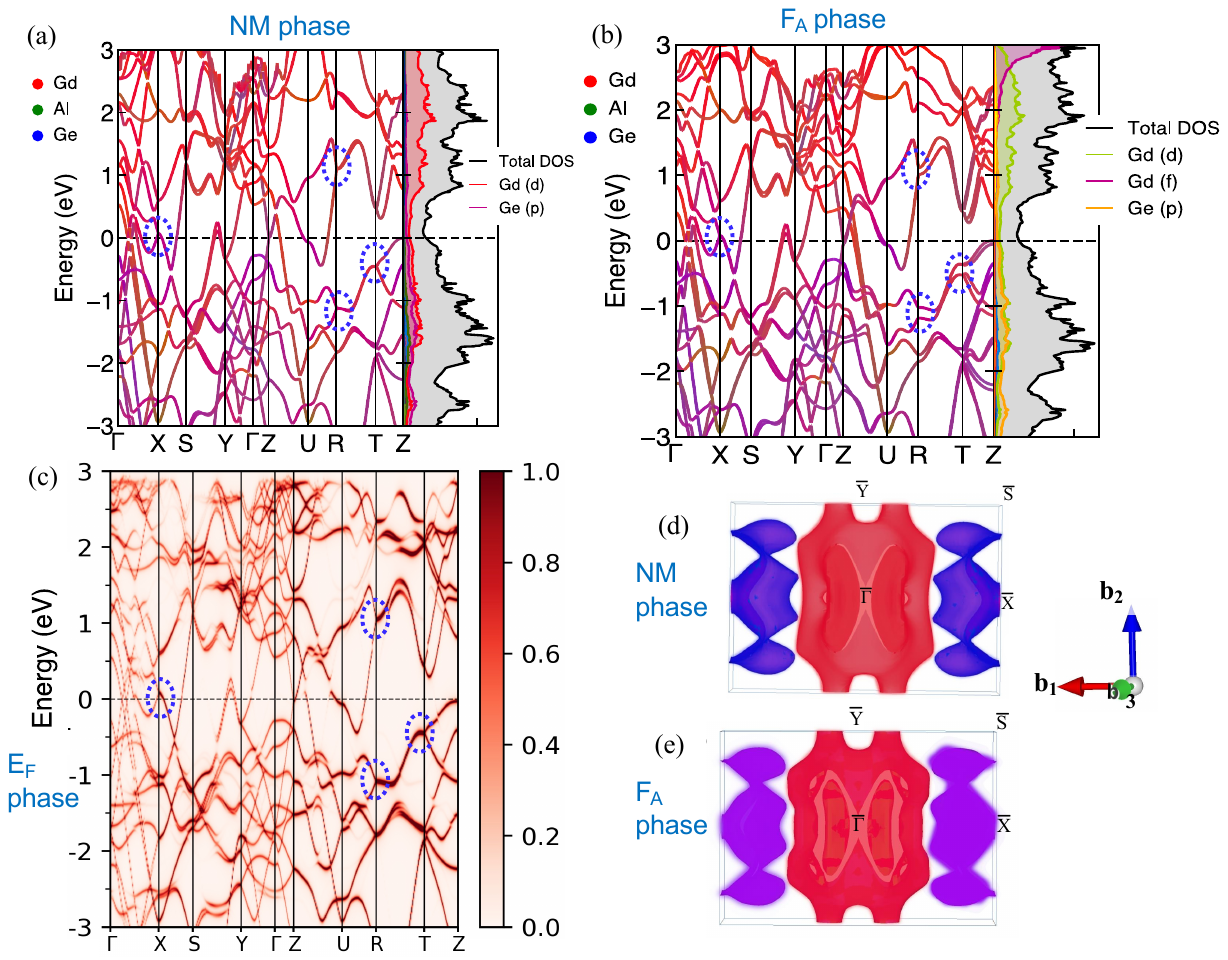}
\caption{Electronic structure of GdAlGe. (a \& b)  Atom resolved electronic band structure and orbital resolved density of states (DOS) with the inclusion of SOC for non-magnetic phase and F$_A$ magnetic phase of GdAlGe, respectively. (c) Unfolded electronic spectra of E$_F$ type AFM pattern. The blue dotted circles in Figs. (a-c) highlight the nodal line features which are partially gapped in F$_A$ phase but remain intact in NM and E$_F$ phases. (d \& e) $k_z$-projected Fermi surface for the NM and F$_A$ phase. The red (blue) colored FS sheets show hole (electron) pockets. These figures show that the electronic structure of NM phase is almost identical to the magnetic phase revealing weak coupling between conduction electron and Gd-moments.}
\label{Fig5}
\end{figure*}

\begin{table}[t]
\begin{center}
\caption{\label{table:MagneticPatternEnergy}Calculated energy difference per formula unit in meV for different magnetic patterns (see text). GGA+SOC+U calculations are done only for the competing configurations, such as E and F-types, where U = 6 eV.}

\label{table:energetics}
\begin{tabular}{p{70pt}|p{30pt}p{55pt}p{55pt}}
\hline\hline
 Pattern & GGA & GGA+SOC & GGA+SOC+U \\ \hline
 E$_F$ & 0 & 0 &  0 \\
 E$_A$ & 3.5 & 3.4 & -\\
F$_A$ & 2.1 & 2.1  & 1.5\\ 
F$_F$ (FM) & 7.9 & 7.7 & -  \\ 
G$_A$ & 6.1 & 6.2  & -\\
 G$_F$ & 5.2 & 5.3  & -\\
C$_A$/C$_F$ & 2.2 & 2.2 & - \\
 \hline\hline
\end{tabular}
\end{center}
\end{table} 

\begin{figure*}
		\begin{center}
		  \includegraphics[width=0.9\textwidth]{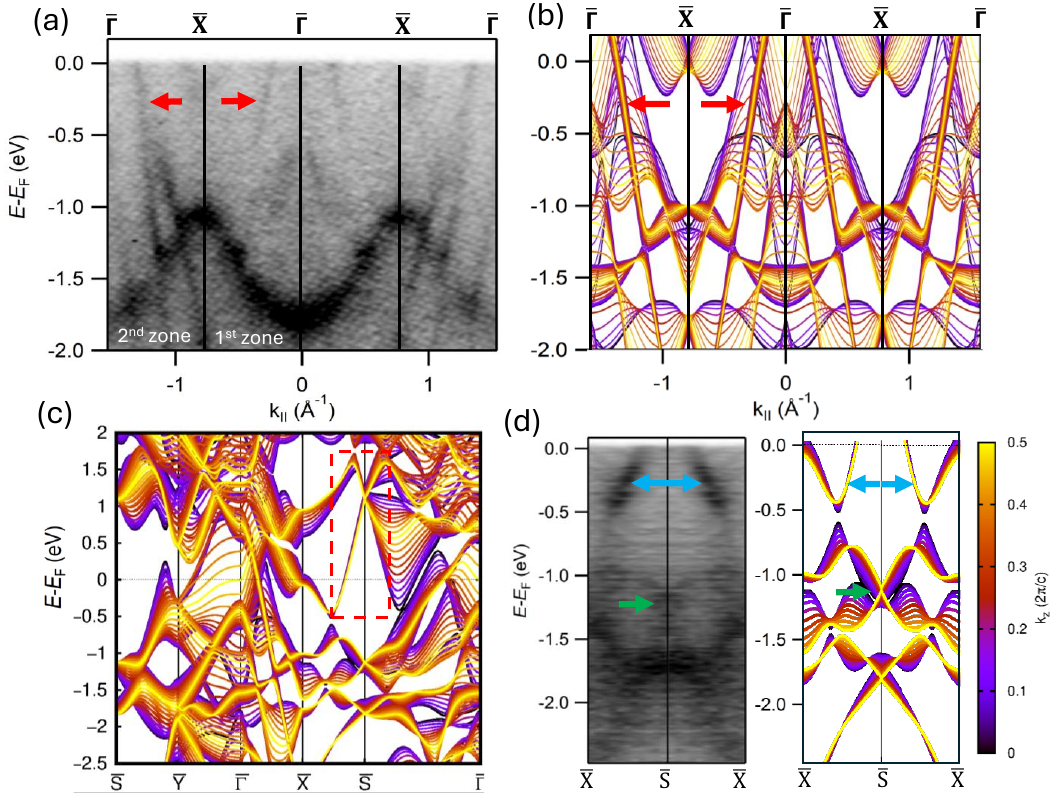}
        \end{center}
		\caption{Comparison of electronic structure between ARPES and DFT (non magnetic). (a) ARPES spectrum at 18 K along the high symmetry path $\bar{\Gamma}$-$\bar{X}$-$\bar{\Gamma}$-$\bar{X}$-$\bar{\Gamma}$ using a photon energy of 200 eV. (b) $k_z$ projected band structure from DFT. Arrows (red) indicate the linear dispersive band crossing the Fermi energy. (c) Multiple Dirac-like band crossing is found at $\bar{S}$ point. The Dirac band with large energy range is marked by a red dashed rectangle. (d) ARPES spectrum (left) and DFT (right) band structure along $\bar{X}$-$\bar{S}$-$\bar{X}$. Arrows highlighting the Dirac-like bands (blue) and Dirac point (green).}
		\label{Fig6}
	\end{figure*}
 
In Fig.~\ref{Fig5}, we compare the DFT calculated electronic structure of GdAlGe between non-magnetic (NM), F$_A$ phase, and E$_F$ magnetic phases. The calculation for the NM phase was done within the open-core approximation where the Gd-4\textit{f} states were frozen. Figure~\ref{Fig5}(a) shows atom-projected electronic bands and density of states which reveal that the states around the Fermi energy are derived mainly from Gd-\textit{d} and Ge-\textit{p} hybridization. 
 The band structure shows metallic character with large Fermi surface and density of states. 
 There are several symmetry-protected multifold band degeneracies at high symmetry points and along high symmetry directions. For example, along the $R$-$T$ direction, the band is eight-fold degenerate, giving rise to a Dirac nodal line-like feature.
 Figure~\ref{Fig5}(b) shows electronic structure for the F$_A$ magnetic pattern. Its electronic structure is very similar to the NM phase except for small exchange splittings that lift band degeneracies, such as at $R$ and $T$ high symmetry points, which are highlighted by blue dotted circles.
 The occupied Gd-\textit{f} states are about 10 eV below the Fermi energy. 
In Fig.~\ref{Fig5}(c), we show the unfolded electronic spectra of E$_F$ type AFM pattern in the BZ of the primitive unit cell for better comparison with Figs.~\ref{Fig5}(a) and (b). The unfolded  electronic spectra is very similar to the previous cases. More importantly, the multi-fold band degeneracies at $R$ and $T$ high symmetry points that were lifted by few meVs in the F$_A$ phase remain intact in the E$_F$ phase, which indicates that nodal-line features are protected in this magnetic phase. The Fermi surface plots for NM and F$_A$ phase  are shown in panels \ref{Fig5}(c) and \ref{Fig5}(d), respectively.  As discussed before, the FS for NM phase is very similar to the mangetic phase. A larger hole-like FS is present around the zone center whereas  electron FS appears at the zone boundary.

 The comparison between ARPES measured and DFT-calculated band structure is shown in Fig.~\ref{Fig6}. Since there are negligible differences between the electronic structures of NM and magnetic phases, for simplicity, we compare ARPES spectra with NM dispersion. In the low-energy region, the ARPES spectrum [Fig.~\ref{Fig6}(a)] shows linearly dispersive hole-like bands (indicated by red arrows) around $\bar{\Gamma}$ that cross the Fermi energy. These bands are more visible in the second Brillouin zone than the first zone, possibly due to the strong matrix element effect. The ARPES spectrum also shows a high-intensity parabolic band with a bottom at $-1.8$~eV at $\bar{\Gamma}$. The $k_z$ projected band structure calculations [Fig.~\ref{Fig6}(b)] along the same path as the ARPES spectrum well captures the features seen in ARPES. Furthermore, DFT calculations show multiple Dirac-like band crossing at $\bar{S}$ point [Fig.~\ref{Fig6}(c)], which are further confirmed by our ARPES results [Fig.~\ref{Fig6}(d)]. Interestingly, most of these bands forming Dirac-like crossings in ARPES, including the linearly dispersing band crossing the Fermi energy, are formed by nodal lines along X-S direction [Figs.~\ref{Fig5}(a-c)]. Overall, most of the features in the ARPES spectrum are well captured by our band structure calculations. It is generally known that the Fermi energy in actual materials may be slightly different from the DFT calculations and a small shift of Fermi energy is sometimes necessary for the best fit between them. As seen from the ARPES vs. DFT comparison, a downward shift of the DFT bands by about 100 meV yields a better comparison with the ARPES spectra. Indeed, when we shift the Fermi energy by about 100 meV, we find that the electronic structure is perfectly compensated, which could also explain the experimentally observed large magnetoresistance.

\section{Discussions and conclusions}
 A notable feature of GdAlGe is the large energy range of the linear band dispersion crossing the Fermi energy {often associated with the Dirac bands. As seen in Fig.~\ref{Fig6}(c), there is a linear band crossing the Fermi energy traversing along the $\bar{X}-\bar{S}$ direction with multi-fold band degeneracies at the $\bar{S}$ point. However, along the $k_x$ direction ($\bar{S}-\bar{Y}$), the band is less dispersive, giving rise to a nodal-line feature. Similarly, a highly dispersive linear band exists along $\bar{\Gamma}-\bar{X} $ direction [Fig.\ref{Fig6}(b)] whereas along the $\bar{\Gamma}-\bar{Y}$ direction, the band is  flatter. In this sense, there is no three-dimensional Dirac cone present in the band structure; thus, we call this band feature Dirac-like. While most of the topological materials studied so far exhibit linear band dispersion up to a few hundred milli-electronvolts from the Dirac point, such as 0.3 eV in Cd$_3$As$_2$ \cite{Cd3AS2_Natcomn_2014}, 0.8 eV in ZrTe$_5$ \cite{ZrTe5_Nature_phys_2016}, 0.8 eV in Na$_3$Bi \cite{Na3Bi_Science_2015}, 0.3 eV in MnBi$_2$Te$_4$ \cite{MnBi2Te4_Nature_2019}, and 0.4 eV in Co$_3$Sn$_2$S$_2$ \cite{Co3Sn2S2_Nature_comn_2018}. In GdAlGe, the range is observed to be as high as $\sim 1.5$ eV in some regions of the Brillouin zone. Such an exceptionally large energy range ($\sim 1.5$ eV) has been observed in a non-magnetic Dirac semimetal ZrSiS \cite{ZrSiS_Sci_adv_2019}, but has not yet been reported in any magnetic topological materials. From the ARPES results, we calculate the Fermi velocity. Surprisingly, we obtain an exceptionally high Fermi velocity of $\sim 6.67$ eV\AA ($\sim c/300$, where $c$ is the speed of light) in GdAlGe, which is comparable with the Fermi velocity reported in other well known Dirac materials such as LaAlGe ($\sim 2.2$ eV\AA) \cite{LaAlGe_APL_2020}, ZrSiS ($\sim 6.7$ eV\AA) \cite{ZrSiS_natcomn_2016} , Cd$_3$As$_2$ ($\sim 9.8$ eV$\AA$) \cite{Cd3AS2_Natcomn_2014} and ZrTe$_5$ ($\sim 6.4$ eV$\AA$) \cite{ZrTe5_Nature_phys_2016}, all of them are non-magnetic. 
 
 To observe the effect of linearly dispersive bands on electronic transport properties in topological semimetals, the primary requirement is that the Fermi energy of the material should remain within the linear dispersion region. In many real materials, the Fermi energy is rather far away from the linear dispersive region. To bring the Fermi energy into the linear dispersive regime, very careful and delicate synthesis procedures including doping are often required. In contrast, a material with a large energy range of linear band dispersion can be robust enough to satisfy this requirement even when the Fermi energy of the crystals is not close to the Dirac/Weyl nodes. 
  
 It is not clear if GdAlGe supports helical or spiral magnetism, which was proposed for GdAlSi \cite{GdAlSi_PRB_2024}, related to a very weak magnetocrystalline anisotropy---a common feature for both. To understand magnetism and its interplay with topological fermions in GdAlGe, it is necessary to explore the nature of magnetic ordering in orthorhombic Dirac nodal line metals, which interestingly has never been investigated thus far. Since Gd is a strong neutron absorber, resonant elastic x-ray scattering (REXS) may be a suitable tool for exploration.

 In summary, we find that replacing Si by Ge in GdAlSi creates sufficient chemical pressure to change the crystal structure from tetragonal, as in GdAlSi, to orthorhombic in GdAlGe. This structural change induced by chemical pressure is unique in the $R$Al$X$ family. Our crystal engineering results in a transformation of electronic state from Weyl semimetal (GdAlSi) to nodal-line metal (GdAlGe). The antiferromagnetic ordering temperature ($T_\mathrm{N1}$ = 35 K) observed in GdAlGe is the highest among $R$Al$X$ family of magnetic topological materials. All the electronic bands that cross Fermi level in GdAlGe are linearly dispersive Dirac-like bands, with the linear dispersion spans over an exceptionally large energy range of $\sim 1.5$ eV along some directions. Dirac/Weyl fermions, characterized by linear energy-momentum dispersion, are topologically protected by chirality, enabling coherent charge transport without dissipation \cite{QL_Chiral_1, Kharzeev2020}. As a result, the electrical conductivity of topological materials shows high robustness toward dimension scaling down to nanoscale, which is a key requirement in selecting energy efficient interconnecting materials for future microelectronics \cite{science.adk6189}. Notably, GdAlSi shows a high and scalable electrical conductivity down to a single monolayer \cite{GdAlSi_monolayer_AM_2025}. The bulk electrical resistivity of GdAlGe is three times lower than that of GdAlSi at room temperature, positioning GdAlGe as an even more attractive material for nano scaled microelectronics applications.
 

\section*{ACKNOWLEDGMENTS}
The research at Brookhaven National Laboratory was supported by the U.S. Department of Energy, Office of Basic Energy Sciences, Contract No. DE-SC0012704. ARPES measurements used resources at 21-ID (ESM) beamline of the National Synchrotron Light Source II, a U.S. Department of Energy (DOE) Office of Science User Facility operated for the DOE Office of Science by Brookhaven National Laboratory under Contract No. DE-SC0012704. This research used Electron Microscopy resources of the Center for Functional Nanomaterials (CFN), which is a U.S. DOE Office of Science User Facility, at Brookhaven National Laboratory under Contract DESC0012704. 

\bibliography{References}	
\end{document}